\documentstyle[12pt,aps]{revtex}
\newcommand{\bold}[1]{\mbox{\boldmath $#1$}}
\tighten
\begin{document}
\draft

\title{\flushleft Comment on `About the magnetic field of a finite 
wire'\footnote{This comment is written by V Hnizdo
in his private capacity. No official support or endorsement by Centers
for Disease Control and Prevention is intended or should be inferred.}} 
\author{\flushleft V Hnizdo}

\address{\flushleft
National Institute for Occupational Safety and Health,\\
1095 Willowdale Road, Morgantown, WV 26505, USA}

\address{\flushleft\rm
{\bf Abstract}. 
A flaw is pointed out in the justification given by  Charitat and Graner 
[2003 {\it Eur. J. Phys.} {\bf 24}, 267]
for the use of the Biot--Savart law
in the calculation of the magnetic field
due to a straight current-carrying wire of finite length.}

\maketitle

\section*{}
\noindent 
Charitat and Graner (CG) \cite{CG} apply the Amp\`ere  and
Biot-Savart laws to the problem of the magnetic field due to
a straight current-carrying wire of finite length, and note that
these laws lead to different results. According to the Amp\`ere law,
the magnitude $B$ of the
magnetic field at a perpendicular distance $r$ from the midpoint along
the length $2l$ of a wire carrying a current $I$ is
$B=\mu_0 I/(2\pi r)$, while the Biot--Savart law gives this quantity as 
$B=\mu_0 Il/(2\pi r\sqrt{r^2+l^2})$
(the right-hand side of equation (3) of \cite{CG} for $B$ cannot be 
correct
as  $\sin\alpha$ on the left-hand side there must equal 
$l/\sqrt{r^2+l^2}$).
To explain the fact that the Amp\`ere and  Biot--Savart laws
lead here to different results, CG say that the former
is applicable only in a time-independent magnetostatic situation, whereas 
the latter
is `a general solution of Maxwell--Amp\`ere equations'. A straight wire of 
finite length
can carry a current only when there is a current source at one 
end of the wire and a curent sink at the other end---and this is possible
only when there are time-dependent net charges at both ends of the wire. 
These charges create a time-dependent electric field and thus the
problem is outside the domain of magnetostatics (we note that the 
time-dependent flux of this electric field is used in equation
(7) of \cite{CG} with the incorrect sign; the desired result for
the magnetic field can be there obtained simply by using
the fact that $\partial q(t)/\partial t=-I$).

We would like to point out that both the Coulomb and Biot--Savart laws 
have to be generalized to give correctly the electric and magnetic 
fields
due to general time-dependent charge and current sources $\rho(\bold{r},t)$ 
and $\bold{J}(\bold{r},t)$. These generalizations are given by 
Jefimenko's equations \cite{OJ,Jack,Griff}:
\begin{equation}
\bold{E}(\bold{r},t)=\frac{1}{4\pi\epsilon_0}\int {\rm d}^3r'
\left[\frac{\rho(\bold{r}',t')}{R^2}\hat{\bold{R}}+\frac{1}{cR}
\frac{\partial\rho(\bold{r}',t')}{\partial t'}\hat{\bold{R}}
-\frac{1}{c^2R}\frac{\partial\bold{J}(\bold{r}',t')}
{\partial t'}\right]_{t'=t-R/c}
\end{equation}
\begin{equation}
\bold{B}(\bold{r},t)=\frac{\mu_0}{4\pi}\int {\rm d}^3r'
\left[\frac{\bold{J}(\bold{r}',t')}{R^2}+\frac{1}{cR}
\frac{\partial\bold{J}(\bold{r}',t')}{\partial t'}\right]
_{t'=t-R/c}{\bold{\times}}\hat{\bold{R}}
\end{equation}
where $R=|\bold{r}-\bold{r}'|$, 
$\hat{\bold{R}}=(\bold{r}-\bold{r}')/R$, and $t'=t-R/c$ is the retarded time.
Equations (1) and (2) are the exact causal (retarded) solutions to Maxwell's
equations for given time-dependent  charge and current
densities $\rho(\bold{r},t)$ and $\bold{J}(\bold{r},t)$.
They can be obtained from the retarded scalar and vector potentials
\begin{equation}
V(\bold{r},t)=\frac{1}{4\pi\epsilon_0}\int {\rm d}^3r'
\frac{\rho(\bold{r}',t-R/c)}{R}\;\;\;\;\;\;
\bold{A}({\bold r},t)=\frac{\mu_0}{4\pi}\int {\rm d}^3r'
\frac{\bold{J}(\bold{r}',t-R/c)}{R}
\end{equation}
of the well-known Lorentz-gauge solution to Maxwell's equations
by performing the spatial differentiations in 
\begin{equation}
\bold{E}(\bold{r},t)=-\bold{\nabla}V(\bold{r},t)
-\partial\bold{A}(\bold{r},t)/\partial t\;\;\;\;\;\;
\bold{B}(\bold{r},t)=\bold{\nabla\times A}(\bold{r},t)
\end{equation}
with due regard for the implicit dependence of the
quantities $\rho(\bold{r}',t{-}R/c)$ and $\bold{J}(\bold{r}',t{-}R/c)$
on the coordinate $\bold{r}$ through the retarded time 
$t-|\bold{r}-\bold{r}'|/c$ \cite{Griff}.
A comparison of equation (3) with equations (1) and (2) shows
that while the Lorentz-gauge potentials can be obtained by a straightforward
`retarding' of the electrostatic scalar and magnetostatic vector
potentials, the general expressions for the fields directly
in terms of the sources contain terms that are additional to those 
of the simply `retarded' Coulomb and Biot--Savart laws.
 
Jefimenko's equation (2) for the magnetic field  reduces to
the standard Biot--Savart law only when one can replace in the integrand
the retarded time $t'$ with the present time $t$ and drop the term with 
the time derivative of the current density. Obviously, this can be done 
when the current density is constant in time, but, nontrivially,
also when the current density at the retarded time $t'$ is given
sufficiently accurately by the first-order Taylor expansion about the present
time $t$ (see \cite{Griff}, problem 10.12):
\begin{equation} 
\bold{J}(\bold{r}',t')\approx\bold{J}(\bold{r}',t)+(t'-t)
\frac{\partial\bold{J}(\bold{r}',t)}{\partial t}
=\bold{J}(\bold{r}',t)-\frac{R}{c}\frac{\partial\bold{J}(\bold{r}',t)}
{\partial t}.
\end{equation}

The Biot--Savart solution to the problem under discussion is thus correct  when
the condition (5) is satisfied, as, for example, when the current $I$ is 
constant in time or varies with  time only linearly. 
However, CG's contention that the standard Biot--Savart law is a general
solution of the Maxwell--Amp\`ere equation 
$\bold{\nabla{\times}B}=\mu_0\bold{J}+c^{-2}\partial\bold{E}/\partial t$
is incorrect. 
Contrary to an assertion of CG, the curl of a non-retarded vector potential
(the $r^2$ in equation (8) of \cite{CG} for this quantity 
is obviously a misprint) does not result in a magnetic field that satisfies  
that equation in general. 
Only the curl of the retarded vector potential, which yields
Jefimenko's generalization (2) of the Biot--Savart law,
\begin{equation}
\frac{\mu_0}{4\pi}\bold{\nabla}\bold{\times}\int {\rm d}^3r'
\frac{\bold{J}(\bold{r}',t{-}R/c)}{R}
=\frac{\mu_0}{4\pi}\int {\rm d}^3r'
\left[\frac{\bold{J}(\bold{r}',t')}{R^2}+\frac{1}{cR}
\frac{\partial\bold{J}(\bold{r}',t')}{\partial t'}\right]
_{t'=t-R/c}{\bold{\times}}\hat{\bold{R}}
\end{equation}
is, together with Jefimenko's generalization (1) of the Coulomb law,
the solution of Maxwell's equations with general time-dependent sources.

Charitat and Graner have expressed an agreement with the substance of this 
comment \cite{CGpersonal}.

\end{document}